\documentclass[prd,twocolumn]{revtex4}
\usepackage{graphicx}
\usepackage{amssymb}

\begin{document}

\title{Big bang nucleosynthesis with a varying fine structure constant and non-standard expansion rate}
\author{Kazuhide Ichikawa}
\author{Masahiro Kawasaki}
\affiliation{Research Center for the Early Universe, University 
  of Tokyo, Bunkyo-ku, Tokyo 113-0033, Japan}
\date{\today}

\begin{abstract}
We calculate primordial abundances of light elements produced during big bang nucleosynthesis when the fine structure constant and/or the cosmic expansion rate take non-standard values. We compare them with the recent values of observed D, $^4$He and $^7$Li abundances, which show slight inconsistency among themselves in the standard big bang nucleosynthesis scenario. This inconsistency is not solved by considering either a varying fine structure constant or a non-standard expansion rate separately but solutions are found by their simultaneous existence. 
\end{abstract}
\maketitle

Big Bang nucleosynthesis (BBN) theory calculates how much light elements are produced in the early universe. The standard BBN takes initial amount of baryons as only one input which is parametrized by baryon number density divided by photon number density: $\eta \equiv n_b/n_{\gamma}$. With $\eta \sim O(10^{-10})$, the theory successfully predicts observed D, $^4$He and $^7$Li abundances extending over ten digits. 

However, the recent measurements of the primordial light elements abundances indicate that the success does not seem to be perfect \cite{Cyburt:2003fe,Steigman:2003gy,Coc:2003ce,Coc:2004ij,Cyburt:2004cq}. Such discrepancy between the observation and the theory is most likely ascribed to the existence of unknown systematic errors in the observation. It is usually considered that systematic errors in D observation are smaller than those in $^4$He and $^7$Li because D is observed in primordial objects, quasar absorption systems, but regression with respect to metallicity is necessary to deduce primordial $^4$He and $^7$Li abundance. In addition, the observed D abundance is consistent with the baryon density from the cosmic microwave background (CMB) data \cite{Spergel:2003cb}, which supports its robustness. Therefore, from this viewpoint, unexplored systematic errors in both $^4$He and $^7$Li measurements solve the discrepancy \footnote{Or, higher $^7$Li value such as measured in Ref.~\cite{Bonifacio:2002yx} may be a solution. However, it will be only marginal due to the discrepancy between D and $^4$He.}.

From another viewpoint, the investigations correctly estimate the systematic errors in $^4$He and $^7$Li measurements so that the discrepancy is solved by non-standard physics. The recent studies on non-standard BBN include non-standard expansion rate (number of neutrino species other than 3) \cite{Barger:2003zg}, lepton asymmetry \cite{Barger:2003rt}, and a varying fine structure constant, $\alpha$ \cite{Nollett:2002da}. They can solve discrepancy between either D and $^4$He or D and $^7$Li but solution for three elements together is not obtained by their individual application \footnote{Ref.~\cite{Dmitriev:2003qq} discusses that a varying deuteron binding energy may have the capacity to render internal agreement between the light element abundances.}.

In this paper, we show the current measurement of the three light elements is consistent without invoking further observational systematic errors if $\alpha$ is higher than today's value during BBN {\it and} the expansion rate is slower than the standard value. One might think that three parameters ($\eta$, $\alpha$ and the expansion rate) necessarily explain any three observations, but  since the combination of non-standard expansion rate and lepton asymmetry (which has been investigated in Ref.~\cite{Barger:2003rt}) can only solve D-$^4$He discrepancy, it is worth searching some combination to reconcile the three. Moreover, varying $\alpha$ and non-standard expansion of the universe may both appear from common models based on string theory which accommodates both a dynamical origin of coupling constants and unusual characteristics of spacetime such as extra dimensions.

\begin{figure}
\includegraphics[width=7cm]{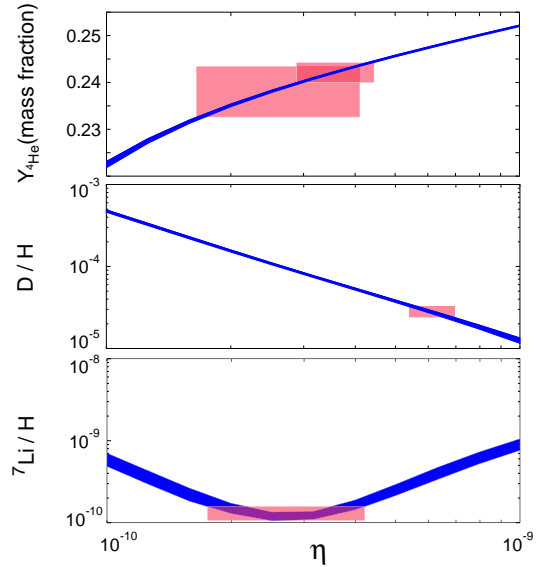}
\caption{Standard BBN calculations of $^4$He, D and $^7$Li abundances as functions of $\eta$ are indicated by three curves whose width shows theoretical 1$\sigma$ uncertainty. The observational 1$\sigma$ uncertainties are expressed by the vertical extension of the boxes. They are drawn to overlap the theory curves so that their horizontal extension shows allowed range of $\eta$. The larger box for $^4$He is from Ref.~\cite{Fields1998} and the smaller from Ref.~\cite{Izotov:2003xn}}
\label{fig:SBBN}
\end{figure}

First of all, we summarize in Fig.~\ref{fig:SBBN}, the current status of predicted and measured abundances of D, $^4$He and $^7$Li. Theoretical uncertainties are computed through Monte Carlo simulations using the values of Ref.~\cite{Cyburt:2001pp} based on the reaction rates of Ref.~\cite{NACRE}. Measured values are taken from Refs.~\cite{Fields1998} (Eq.~(\ref{eq:obs_He_FO})) and \cite{Izotov:2003xn} (Eq.~(\ref{eq:obs_He_IT})) for $^4$He, from Ref.~\cite{Kirkman:2003uv} for D, and from Ref.~\cite{Ryan2000} for $^7$Li: 
\begin{eqnarray}
{\rm Y_{^4He,FO}} &=& 0.238 \pm 0.002 \pm 0.005, \label{eq:obs_He_FO}\\
{\rm Y_{^4He,IT}} &=& 0.2421 \pm 0.0021, \label{eq:obs_He_IT}\\
{\rm (D/H)} &=& 2.78^{+0.44}_{-0.38} \times 10^{-5}, \label{eq:obs_D}\\
{\rm (^7Li/H)} &=& 1.23^{+0.68}_{-0.32} \times 10^{-10}\ (95\%), \label{eq:obs_Li_R}
\end{eqnarray}
In Eq.~(\ref{eq:obs_He_FO}), the first uncertainty is statistical and the second one is systematic. Their root-mean-square, $[{\rm (stat.)}^2 + {\rm (syst.)}^2]^{1/2}$, is the combined 1$\sigma$ error. For asymmetric errors, we adopt conservatively the larger one as 1$\sigma$ error (for $^7$Li, we divide the error in Eq.~(\ref{eq:obs_Li_R}) by 2 to make it 1$\sigma$). From the figure, we see $^4$He and $^7$Li are compatible with $\eta \approx (2\sim 4)\times 10^{-10}$ but higher baryon density $\eta \approx 6 \times 10^{-10}$ is necessary for D. On performing $\chi^2$ analysis, due to the severer D-$^7$Li discrepancy, we do not have a $\eta$ range to explain three elements abundances together with standard BBN at 99\% confidence level (for either $^4$He observations). We stress once again that such a discrepancy first requires a reassessment of systematic effects in the measurements of primordial abundances (especially that of $^7$Li), but below, assuming further systematic errors are not found, we investigate whether this discrepancy is solved by considering varying $\alpha$ and/or non-standard expansion rate.

\begin{figure}
\includegraphics[width=7cm]{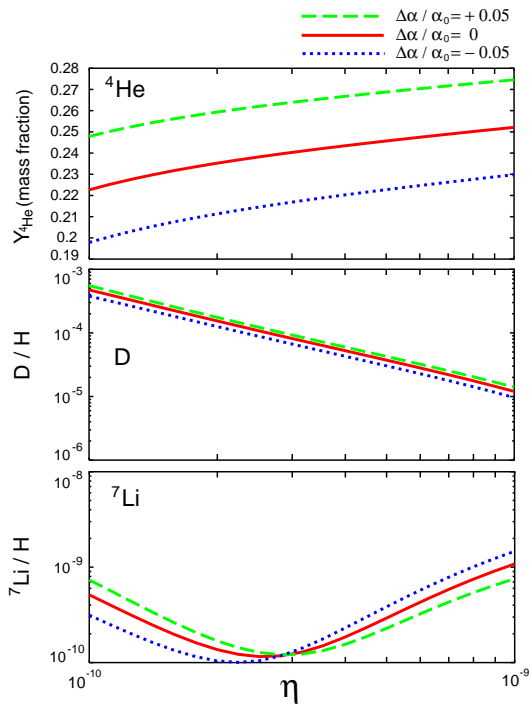}
\caption{$\alpha$-dependence of light element abundances. The cases for $\Delta \alpha/\alpha_0=(\alpha-\alpha_0)/\alpha_0=0$, $+0.05$, $-0.05$ are drawn with solid, dashed,  and dotted lines.}
\label{fig:varyalpha}
\end{figure}

We calculate BBN abundances when $\alpha$ is different from the measured value at present, $\alpha_0 \approx 1/137$, as is described in our previous paper \cite{Ichikawa:2002bt} \footnote{Here we only vary $\alpha$ and do not consider possible change in QCD scale, $\Lambda_{QCD}$, of which variation was taken into account in Ref.~\cite{Ichikawa:2002bt}. The relation between $\Delta\alpha$ and $\Delta\Lambda_{QCD}$ is model dependent and it is possible that $\Delta \alpha/\alpha \gg \Delta \Lambda_{QCD}/\Lambda_{QCD}$ \cite{Dine:2002ir}.}. We have added the Ref.~\cite{Nollett:2002da}'s improvements concerning $\alpha$-dependence of the nuclear reaction rates and binding energies. Fig.~\ref{fig:varyalpha} shows how abundances change when $\alpha$ is varied, reproducing the results of Ref.~\cite{Nollett:2002da}. The dependence is understood as follows. For $^4$He, since increasing $\alpha$ decreases the neutron-proton mass difference (a proton is electrically charged so it becomes heavier than a neutron, which is electrically neutral),  $\Delta m$, the freeze-out ratio of neutron to proton increases and so does $^4$He abundance \cite{EarlyUniverse,Uzan:2002vq}. Meanwhile, other light elements abundances are affected mainly by the change in the Coulomb barrier penetrability for the charged-particle induced nuclear reaction rates \cite{Bergstrom:1999wm,Nollett:2002da}, which is the exponential factor in the following expression of the cross section $\sigma(E)$ at energy $E$,
\begin{eqnarray}
\sigma(E) = \frac{S(E)}{E} \exp \left( -2\pi \alpha Z_i Z_j \sqrt{\frac{\mu}{2E}} \right), \label{eq:crosssection}
\end{eqnarray}
where $S(E)$ is the astronomical $S$-factor, $\mu$ is the reduced mass, and $Z_{i,j}$ are the atomic number of the colliding nuclei. Since larger $\alpha$ suppresses the charged-particle reaction rates, the nucleosynthesis proceeds slower and this saves more D to be burned out. The same is true for T and more of it survives with higher $\alpha$. This explains the $^7$Li increase for lower $\eta$ since $^7$Li is mainly produced by $^4$He(T,$\gamma$)$^7$Li. For higher $\eta$, $^7$Li comes from the electron capture of $^7$Be. $^7$Be is in turn produced through $^4$He($^3$He,$\gamma$)$^7$Be, which is strongly suppressed by higher $\alpha$ because of the large Coulomb barrier  ($Z_i = Z_j = 2$ in Eq.~(\ref{eq:crosssection}) ). 

Especially, the dependence of $^4$He on $\alpha$ is derived from a number ratio of neutron to proton when their interchange freezes out, that is, when the weak interaction becomes comparable to the expansion rate of the universe. Since almost every neutron is synthesized into $^4$He, its mass fraction is approximately expressed by neutron and proton number density, $n_n$ and $n_p$ at freeze-out (denoted by subscript "$f$") as
\begin{eqnarray}
Y_{^4{\rm He}} = \frac{2n_n}{n_n+n_p} \bigg|_f = \frac{2}{1+(n_p/n_n)_f} =\frac{2}{1+e^{\Delta m/T_f}}, \label{eq:freezeout}
\end{eqnarray}
where freeze-out temperature $T_f$ is about $0.7$ MeV. Since $\Delta m$ is measured to be $1.293$ MeV and its electromagnetic part is $-0.76$ MeV \cite{Gasser:1982ap},
\begin{eqnarray}
\Delta m = -0.76 \frac{\alpha}{\alpha_0} + 2.05\ {\rm MeV}. \label{eq:deltam}
\end{eqnarray}
Then $\Delta Y_{^4He}/Y_{^4He} \approx \Delta \alpha/\alpha_0$ follows.

\begin{figure}
\includegraphics[width=7cm]{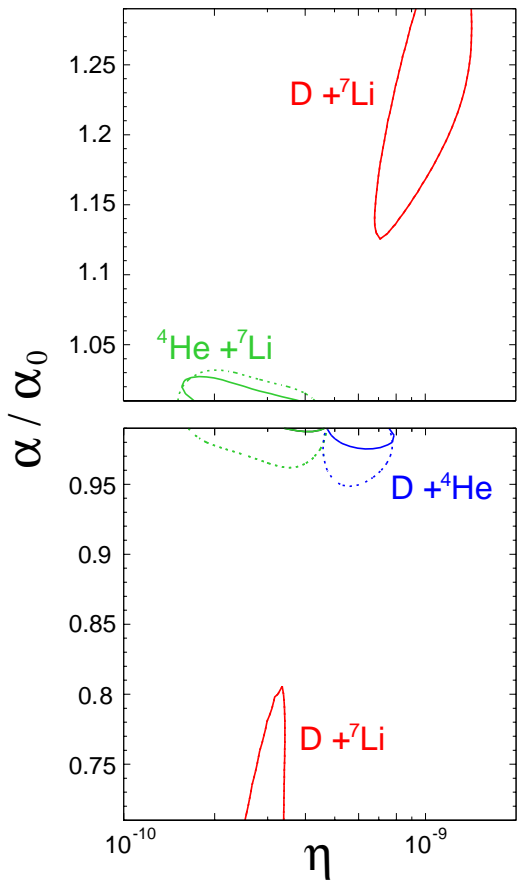}
\caption{Contours show 95\% confidence regions allowed by combinations of two elements observations, D and $^4$He, D and $^7$Li, and $^4$He and $^7$Li. As for combinations including $^4$He, solid lines use $^4$He data of Ref.~\cite{Izotov:2003xn} and dotted lines use those of Ref.~\cite{Fields1998}. A region consistent with three elements abundances together is not found. We see D and $^7$Li are only reconciled by adopting $\alpha \neq \alpha_0$ but that makes $^4$He abundance inconsistent with the observation.}
\label{fig:N=3}
\end{figure}

To figure out whether varying $\alpha$ can solve the discrepancy between D and $^4$He and/or $^7$Li, we calculate $\chi^2$ as a function of $\eta$ and $\alpha$ and search parameter space allowed by the observation of the light elements. The results are summarized in Fig.~\ref{fig:N=3}. We note that we take into account the uncertainty in the present value of the electromagnetic part of $\Delta m$ (which we neglected in Ref.~\cite{Ichikawa:2002bt}). Since Ref.~\cite{Gasser:1982ap} have reported it to be less than $0.3$ MeV, we regard this value to be $3\sigma$ of a gaussian error profile and incorporate in our Monte Carlo simulation along with uncertainties in the reaction rates. This uncertainty does not exist when $\alpha=\alpha_0$, at which theoretical errors become discontinuous and we have to split the $\chi^2$ calculation like Fig.~\ref{fig:N=3}.

\begin{figure}
\includegraphics[width=7cm]{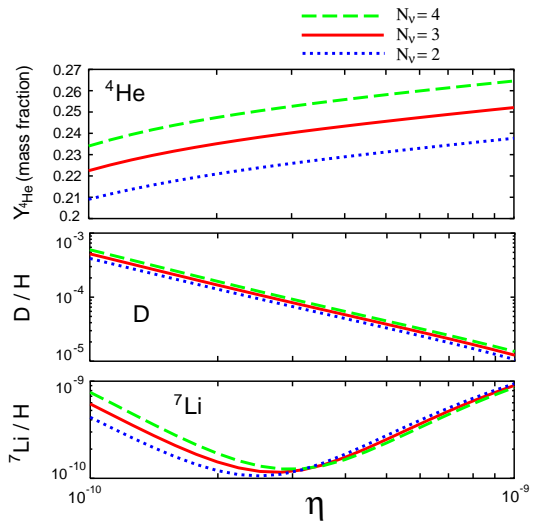}
\caption{$N_{\nu}$-dependence of light elements abundances. The cases for $N_{\nu}=3$, $4$, $2$ are drawn with solid, dashed,  and dotted lines.}
\label{fig:varyN}
\end{figure}

In Fig.~\ref{fig:N=3}, as is expected from Figs~\ref{fig:SBBN} and \ref{fig:varyalpha} or has been demonstrated in Ref.~\cite{Nollett:2002da}, there is no region that explains measurements of the three elements together. Roughly speaking, $^4$He observation constrains in $\alpha$-direction while D and $^7$Li constrain $\eta$-direction because they are more sensitive to corresponding parameters (see Fig~\ref{fig:varyalpha}). The $^4$He$+^7$Li contour lies around $\alpha = \alpha_0$ because they are already consistent in the standard BBN. The D$+^4$He contour lies in $\alpha < \alpha_0$ because, as can be seen in Fig.~\ref{fig:SBBN}, $^4$He is too much synthesized  around the $\eta$ range determined by D and cutting it down by decreasing $\alpha$ makes it consistent with the observation. At last, the D$+^7$Li contours exist in both $\alpha > \alpha_0$ and $\alpha < \alpha_0$. The former corresponds to the region where D determines the $\eta$-range (in which $^7$Li is over-synthesized) and the $\alpha$-direction is constrained by required increase in $\alpha$ to bring down the $^7$Li. Meanwhile, for the latter, since $^7$Li fix the $\eta$-range, $\alpha<\alpha_0$ is necessary to decrease over-produced D. However, such $\alpha\neq\alpha_0$ capable of reconciling D and $^7$Li either over- or under-produce $^4$He.

The story so far on BBN with varying $\alpha$ (and with the observations we adopt) is summarized as follows. It fails to explain the observed three light elements abundances because $\alpha$ required to make D and $^7$Li compatible creates too much or too small $^4$He. Since it is difficult to come up with any non-standard BBN other than varying $\alpha$ which reconciles D and $^7$Li, considering one which recovers $^4$He without violating the success of the varied $\alpha$ on D and $^7$Li would be next best option.

Our choice in this paper is non-standard expansion rate. It is usually treated and parameterized as effective number of neutrino species, $N_{\nu}$, and we follow this convention. The standard BBN corresponds to $N_{\nu}=3$. The dependence of the light elements abundances on $N_{\nu}$ is shown in Fig.~\ref{fig:varyN}, which we explain briefly.  $^4$He is again determined by Eq.~(\ref{eq:freezeout}). Since increasing $N_{\nu}$ means increasing expansion rate, it raises freeze-out temperature and leaves more neutrons which synthesize into $^4$He. Faster expansion rate also makes nucleosynthesis less effective so more D is left unburned. So is T which fuses with $^4$He to be $^7$Li for lower $\eta$. As for $^7$Be (which is the origin of $^7$Li for higher $\eta$), its destruction process $^7$Be(n,p)$^7$Li is enhanced due to increased $n$. The reason for increase in $n$ is same as D.

We see from the figure that $N_{\nu}$ does worse than $\alpha$ as for reconciling D and $^7$Li because $N_{\nu}$ barely alter them (especially for higher $\eta$) while 
changing the $^4$He abundance substantially. However, this turns out to be the advantage when combined with varying $\alpha$ because that is the very required property mentioned above. For $\alpha>\alpha_0$, since $^4$He is overproduced in order to adjust D and $^7$Li, $N_{\nu}$ has to decrease. On the other hand, for $\alpha<\alpha_0$, $N_{\nu}$ has to increase.

\begin{figure}
\includegraphics[width=7cm]{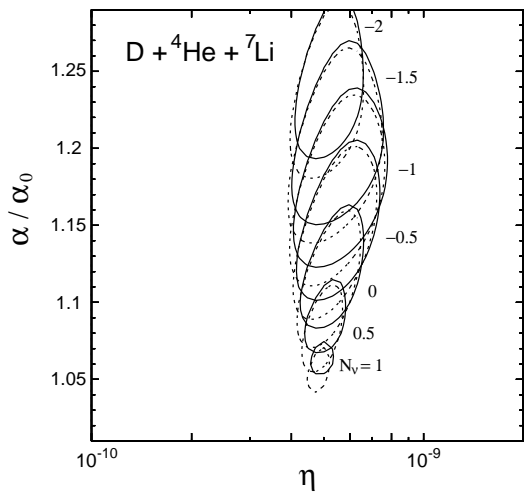}
\caption{Contours show 95\% confidence regions allowed by combinations of all three elements observations, D, $^4$He and $^7$Li, for various $N_{\nu}$. Solid lines use $^4$He data of Ref.~\cite{Izotov:2003xn} and dotted lines use those of Ref.~\cite{Fields1998}. We do not find an allowed region for $\alpha < \alpha_0$ by varying $N_{\nu}$. }
\label{fig:varyalphaN}
\end{figure}

With this insight, we perform $\chi^2$ calculations similar to Fig.~\ref{fig:N=3}, making $N_{\nu}$ less than 3 for $\alpha>\alpha_0$ and more than 3 for $\alpha<\alpha_0$. Naive expectation is that the contours including $^4$He seen in Fig.~\ref{fig:N=3} would approach D$+^7$Li contours and eventually they would merge to form allowed regions of three elements observations for both $\alpha>\alpha_0$ and $\alpha<\alpha_0$. However, while solutions are obtained for $\alpha>\alpha_0$ as Fig.~\ref{fig:varyalphaN}, there is no solution for $\alpha<\alpha_0$. Notice that we look for solutions in the range $0.71 < \alpha < 1.29$, where the modification to the reaction rates caused by varying $\alpha$ is considered to be valid \cite{Nollett:2002da}. The different behavior stems from subtle effects of varying $N_{\nu}$ on D and $^7$Li. With $N_{\nu}<3$, from Figs.~\ref{fig:SBBN} and \ref{fig:varyN}, we see that the measured D and $^7$Li are more consistent (with smaller $\eta$) than the standard BBN because D is predicted to be smaller ($^7$Li is predicted to have larger abundance which means working in opposite way but $N_{\nu}$- and $\eta$-dependences are both larger for D, especially when the measured value of $^7$Li is around the trough of the theoretical curve, so D is thought to be the decisive factor). On the contrary, since $N_{\nu}>3$ makes D larger, higher $\eta$ is more consistent. This requires very small $\alpha$ to decrease D, so small as to be outside the region with theoretical reliability.

Fig.~\ref{fig:varyalphaN} shows our solutions to the inconsistency between the standard BBN and the measured primordial abundances of the light elements, Eqs.~(\ref{eq:obs_He_FO}) to (\ref{eq:obs_Li_R}). The different $^4$He measurements give similar results because the theoretical uncertainty is comparable to the combined uncertainty in the observation of Eq.~(\ref{eq:obs_He_FO}). This mainly comes from the uncertainty in the electromagnetic part of the neutron-proton mass difference because the effect of uncertainties in the reaction rates on $^4$He yield is negligible. The estimation for this error affects the size of the allowed regions but central values and qualitative features do not change. It is concluded that larger $\alpha$ and slower expansion rate (expressed by $N_{\nu}<3$) solve the discrepancy between the standard theory and the observations. The solution with maximum $N_{\nu}$ is found at about $N_{\nu}\approx 1.16$, $\eta \approx 4.7\times 10^{-10}$ and $\alpha/\alpha_0 \approx 1.05$.

As we finish, we would like to make some comments. The first is on possible origins of lower-than-standard expansion rate ($N_{\nu}<3$). We present here three possibilities: 1) non-thermal distribution of active neutrinos caused by low reheating temperature \cite{Kawasaki:1999na,Kawasaki:2000en}; 2) negative dark radiation possibly exists in brane world scenarios \cite{Shiromizu:1999wj,Mukohyama:1999qx}; and 3) a varying (smaller at BBN) gravitational constant which is often found when the Ricci scalar is non-minimally coupled to a scalar field \cite{Uzan:2002vq}. The 1) is limited to $N_{\nu}>0$ but 2) and 3) can be any value as far as the total energy is positive. Their connection to varying $\alpha$ would be promising and quite interesting. The second is consistency with the CMB data. To our knowledge, there is no analysis of CMB data concerning simultaneous change in $\alpha$ and $N_{\nu}$.  Since full statistical treatment is beyond the scope of this paper, we just check the effect of our solution on the first peak. An increase in $\alpha$ raises the first peak \cite{Martins:2003pe} and a decrease in $N_{\nu}$ lowers it \cite{Hu:1995fq}. It is reassuring that some cancellation is likely to take place but details remain to be worked out. Third, another solution should be found by considering lepton asymmetry instead of the non-standard expansion rate because its existence change $^4$He abundance considerably while leaving D and $^7$Li almost unchanged. These issues are discussed in other places \cite{IK}.

 \end{document}